\documentclass[%
 reprint,
superscriptaddress,
 amsmath,amssymb,
 aps,
prb,
]{revtex4-1}

\usepackage{graphicx}
\usepackage{dcolumn}
\usepackage{bm}
\usepackage{amsmath}
\usepackage{gensymb}
\usepackage{natbib}
\usepackage[english]{babel} 
\usepackage{color}

\newcommand{\mf}{Mn$_{1-x}$Fe$_{x}$PSe$_3$}
\newcommand{\m}{MnPSe$_3$}
\newcommand{\f}{FePSe$_3$}
\newcommand{\Mn}{Mn$^{2+}$}
\newcommand{\Fe}{Fe$^{2+}$}

\newcommand{\nn}{\nonumber}

\begin{document}

\title{Strong anisotropy in the mixed antiferromagnetic system Mn$_{1-x}$Fe$_{x}$PSe$_3$}

\author{Ankita Bhutani}
\affiliation{Materials Science and Engineering Department and Materials Research Laboratory, University of Illinois at Urbana-Champaign, Urbana, Illinois, USA.}
\author{Julia L. Zuo}
\affiliation{Materials Science and Engineering Department and Materials Research Laboratory, University of Illinois at Urbana-Champaign, Urbana, Illinois, USA.}
\author{Rebecca D. McAuliffe}
\affiliation{Materials Science and Engineering Department and Materials Research Laboratory, University of Illinois at Urbana-Champaign, Urbana, Illinois, USA.}
\author{Clarina R. dela Cruz}
\affiliation{Neutron Scattering Division, Oak Ridge National Laboratory, Oak Ridge, TN 37831, United States}
\author{Daniel P. Shoemaker}
\affiliation{Materials Science and Engineering Department and Materials Research Laboratory, University of Illinois at Urbana-Champaign, Urbana, Illinois, USA.}
\date{\today}

\begin{abstract}

We report the magnetic phase diagram of Mn$_{1-x}$Fe$_{x}$PSe$_3$ which represents a random magnet system of two antiferromagnetic systems with mixed spin, mixed spin anisotropies, mixed nearest neighbor magnetic interactions and mixed periodicities in their respective antiferromagnetic structure. Bulk samples of Mn$_{1-x}$Fe$_{x}$PSe$_3$ have been prepared and characterized phase pure by powder X-ray and neutron diffraction and X-ray fluorescence. Nature and extent of magnetically ordered state has been established using powder neutron diffraction, dc magnetic susceptibility and heat capacity. Long-range magnetic ordering exists between $x = 0.0$ and 0.25 (MnPSe$_3$-type) and between $x = 0.875$ and $1$ (FePSe$_3$-type). A short-range magnetic order with existence of both MnPSe$_3$- and FePSe$_3$-type nano-clusters has been established between $x = 0.25$ and $0.875$. Irreversibility in dc magnetization measurements, also characterized by isothermal and thermoremanent magnetization measurements suggest similarities to magnetic nanoparticles where uncompensated surface spins result in {\color{black} diverging thermoremanent and isothermal remanent magnetization responses}, further reinforcing existence of magnetic nano-clusters or domains. A spin glass state, observed in analogous Mn$_{1-x}$Fe$_x$PS$_3$, has been ruled out and formation of nano-clusters exhibiting both ordering types results from unusually high anisotropy values. The effect of ligand contributions to the spin-orbit interactions has been suggested as a possible explanation for high $D$ values in these compounds.

\end{abstract}

\pacs{Valid PACS appear here}
\maketitle


\setlength{\parindent}{0.1 in}
\section{Introduction}

%
%
%

Disrupting  the long-range ordering of magnetic systems can
manifest a variety of behaviors in crystalline materials, 
perhaps most notably in the form of emergent properties
such as unconventional superconductivity in iron-based
and cuprate materials.
In those cases, the spin interactions are complex, with a mixture
of local and
itinerant moments and quantum fluctuations, respectively,
leading to complex behavior. 
The superconducting parent compounds could be contrasted with
materials where the behavior is more pedestrian, such as strongly
classical systems where spin-glass behavior arises as multiple
competing order parameters lead to a frozen state.
A third, uncommon scenario can occur when the local coupling is 
strong enough to preclude the spin glass state, and competition
can lead to uncompensated moments via complex domain formation. 
%

A detailed mean-field and renormalization-group study
of the possible magnetic orderings of randomly-mixed magnets
was conducting by Fishman and Aharony in 1978.\cite{fishman1978phase,fishman1979phase,fishman1980phase}
A random magnet containing a mixture of ions with competing 
spin anisotropies orders in a ``mixed phase" or
``oblique antiferromagnetic phase" at intermediate
compositions and the phase diagram of such a magnet 
exhibits a tetracritical ``decoupled" point. Experimental evidence 
of such phases has been observed in the solid-solution
intermetallic Tb$_x$Er$_{1-x}$Ni$_5$ and ionic Fe$_{1-x}$Co$_x$Cl$_2$.\cite{pirogov2009tb,PhysRevB.34.1864} 
On the other hand, mixtures of antiferromagnets with different 
periodicities can form an intermediate phase with both magnetic
orderings, as observed in Fe$_{1-x}$Mn$_x$WO$_4$.\cite{wegner1973magnetic} 
A random magnet with competing interactions forms 
a disordered or spin glass state as observed in Mn$_{1-x}$Fe$_x$PS$_3$.\cite{takano2003magnetic}

Fe$_{1-x}$Mn$_x$WO$_4$ displays a very rich magnetic phase diagram where MnWO$_4$ exhibits 3 types of antiferromagnetic ordering and FeWO$_4$ exhibits only 1 type. A solid solution between the two results in competition between and a coexistence of interpenetrating magnetic structures related to the pure systems MnWO$_4$ and FeWO$_4$.  

\begin{figure}
	\centering\includegraphics[width=0.9\columnwidth]{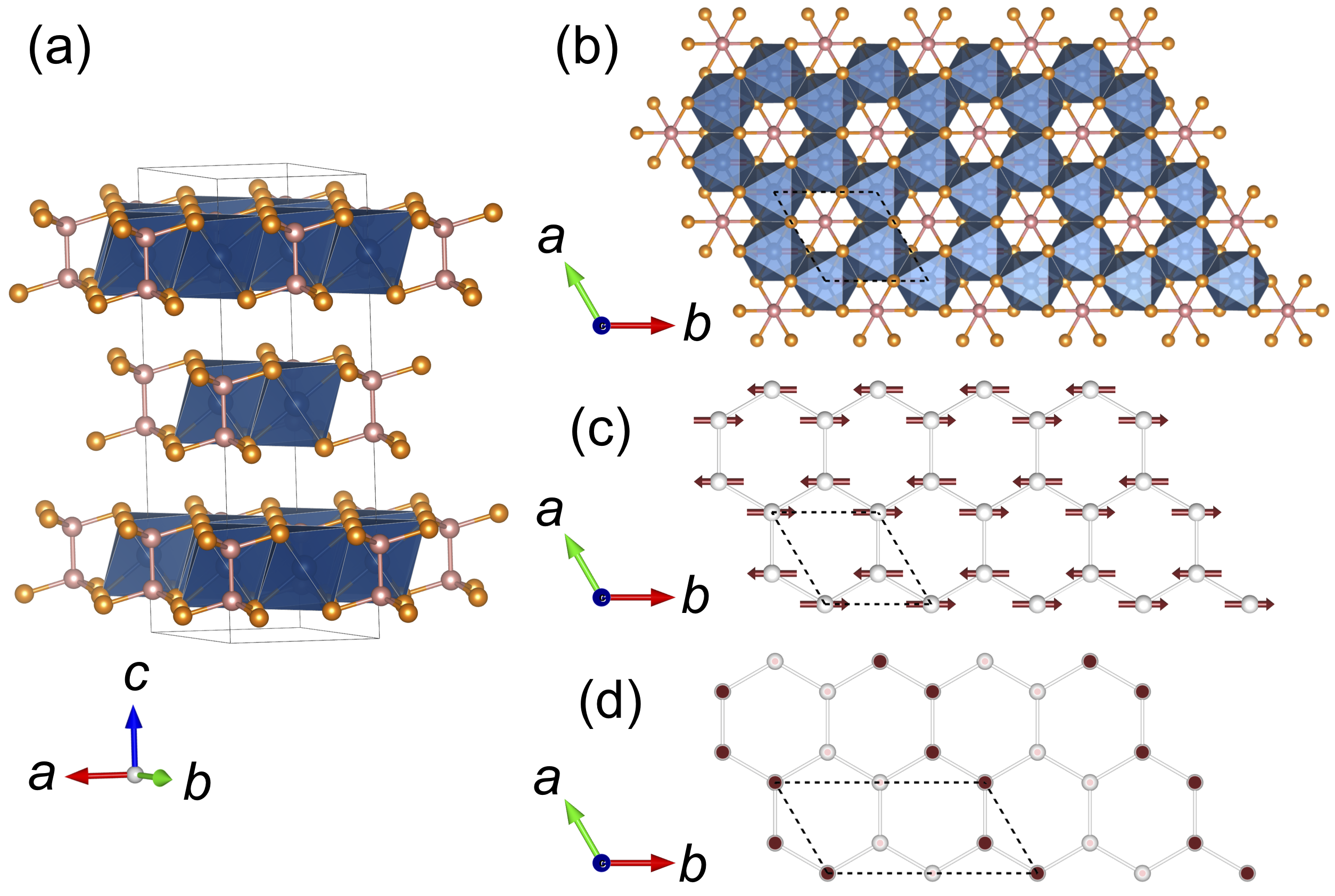} \\
	\caption{{\color{black} (a) Atomic structure of $M$PSe$_3$ for $M$ = Mn or Fe, with transition metals in dark blue octahedra, P (pink) and Se (orange). In (b), a single layer of the structure is shown with the $k=0$ magnetic cell of MnPSe$_3$ dashed. The honeycomb pattern arises from periodic P$_2$Se$_6$ polyanions, where pairs of P are eclipsed in this view. The magnetic structure of \m\ is shown in (c), with \Mn\ only in white. Spins are in the $ab$ plane. In (d), the magnetic structure of \f\ is shown, with spins pointing in the $\pm c$ direction. (b-d) are shown at the same scale. } The direction of the \Mn\ moments in the basal plane was recently found to be canted 8$^\circ$ from $b$.\cite{ressouche2010magnetoelectric}}
	\label{fig:unitcells}
\end{figure} 

Two such compounds that exhibit different magnetic interactions and orderings are \m\ and \f\ belonging to the family of metal thio(seleno)phosphates (MTPs), which are two-dimensional layered compounds with layers bound by weak van der Waals forces. MTPs form a unique family of compounds in which the spin dimensionality may be varied by the choice of the transition metal ion. The MTPs were first discovered by Friedel in 1894.\cite{friedel1894thiohypophosphates} \m\ and \f\ are isostructural and crystallize in the $R\bar{3}$ space group.
{ \color{black} 
	M$_2$P$_2$Se$_6$ can be visualized as $ABCABC$-stacked slabs of CdI$_2$-like units  with 2/3 of the edge-sharing octahedral centers occupied by the transition metal cations, forming a honeycomb network, and the remaining 1/3 occupied by the P--P dimers as shown in Figure \ref{fig:unitcells}. P--P dimers covalently bond to six Se atoms to form (P$_2$Se$_6$)$^{-4}$ ethane-like polyanion units.
}

The magnetic structures for \m\ and \f\ were first examined in 1981 
using neutron powder diffraction by Wiedenmann, \emph{et al.}\cite{wiedenmann1981neutron}
\m\ and \f\ both order antiferromagnetically with $T_N$ of 74 and 119~K and 
Ne\'{e}l vectors  $k = [000]$ and $k = [1/2$ $0$ $1/2]$, respectively.
{ \color{black}
	Layers of both magnetic structures are plotted in Figure \ref{fig:unitcells}(c,d).
}
The magnetic moments of Mn$^{2+}$ ($S = 5/2$) lie in the basal plane 
all three intralayer $J1$ (n), $J2$ (nn) and $J3$ (nnn) interactions are antiferromagnetic. 
On the other hand, the magnetic moments of \Fe\ ($S = 2$) lie along \textit{c}-axis with $J1$ being ferromagnetic, and $J2$ and $J3$ being antiferromagnetic.
\m\ and \f\ can thus be represented as Heisenberg $XY$ and Ising systems, respectively.
A solid solution between \m\ and \f\ thus represents a quite complex 
random alloy, where $S$, $J$, $D$ and $k$ are all competing. 
Such a competition can result in presence of one or more of the theoretically predicted and experimentally realized magnetically ordered phases depending on the chemical composition. Magnetic ordering can, therefore, either be glassy in case of strong competing exchange interactions as observed in sulfides, or be a competing two-phase ordered state in case of strong anisotropic contributions to the total Hamiltonian. 


In this article, we present a detailed investigation of the 
magnetic phase diagram of \mf\ by means of X-ray diffraction, 
X-ray Fluorescence, powder neutron diffraction, DC magnetization 
and heat capacity measurements. Our investigation reveals presence 
of the two end-member magnetic orderings along with a region of 
competing antiferromagnetic orders that exhibits uncompensated moments
and nanoscale domains, as evidenced by broad magnetic diffraction peaks,
despite sharp structural Bragg peaks.

%
%

\section{Experimental Procedure} 

Bulk synthesis of the samples in the solid solution range of Mn$_{1-x}$Fe$_{x}$PSe$_3$ ($0 \leq x \leq 1$, in increments of 0.125) was carried out using traditional solid state synthesis. 
{\color{black}
	Reagents of 
	Mn (crushed granules, Alfa Aesar, 99.98\%), 
	Fe (200 mesh, Alfa Aesar, 99\%), 
	P (red, powder, Sigma-Aldrich, 99.99\%), 
	and Se (crushed granules, Alfa Aesar, 99.999\%) 
	were ground together in an Ar-filled glove box.
}
Precursors were loaded in 12 mm diameter fused silica tubes and sealed under vacuum using liquid nitrogen to prevent P and Se loss during vacuum sealing, and reacted at 650$^\circ$C with a ramp rate of 10$^\circ$C per minute and 30~days hold time, followed by furnace cooling.
{\color{black}
Heating at higher temperatures led to decomposition of the product, and no large crystals were obtained.}

Powder X-ray diffraction measurements were conducted in transmission with a Bruker D8 diffractometer with Mo-K$\alpha$ radiation. Rietveld analysis was carried out using TOPAS 5.~\cite{coelho2004topas}
XRF data were collected using a Shimadzu EDX-7000 spectrometer under a He atmosphere. Three sets of data were collected and averaged
to determine the composition. 

Neutron diffraction data were collected between 1.5~K and 300~K using the HB-2A powder diffractometer at the High Flux Isotope Reactor at Oak Ridge National Laboratory for $x = 0, 0.25, 0.375, 0.5, 0.625$ and $1$. Powders (1-2 g) were loaded in V cans with He exchange gas and measured with incident neutrons with wavelength $\lambda = 2.41$~\AA. Rietveld analyses and magnetic structure solutions were performed with FullProf and SARAh. \cite{rodriguez1990fullprof,wills2000new} 

Magnetic susceptibility measurements were collected on a Quantum Design MPMS 3  magnetometer. Thermoremanent magnetization(TRM) and isothermal remanent magnetization (IRM) measurements were also collected on a Quantum Design MPMS 3  magnetometer. 

The samples were field-cooled
to 5~K, the temperature was stabilized for 10~min, 
field was turned off and the remanent moment was measured at the varying fields.
For IRM measurements, the samples were cooled in zero field to 5~K,
the temperature was stabilized for 10~min, a magnetic field was applied for 10~min and switched off, and remanent magnetic moment was measured. 
Heat capacity measurements were performed using a Quantum Design Dynacool PPMS (Physical Property Measurement System), with pressed pellets mounted using N-grease and a two-tau procedure.

\section{Results and Discussion}

\subsection{Evaluating structure and long-range order}

\begin{figure}
	\centering\includegraphics[width=1\columnwidth]{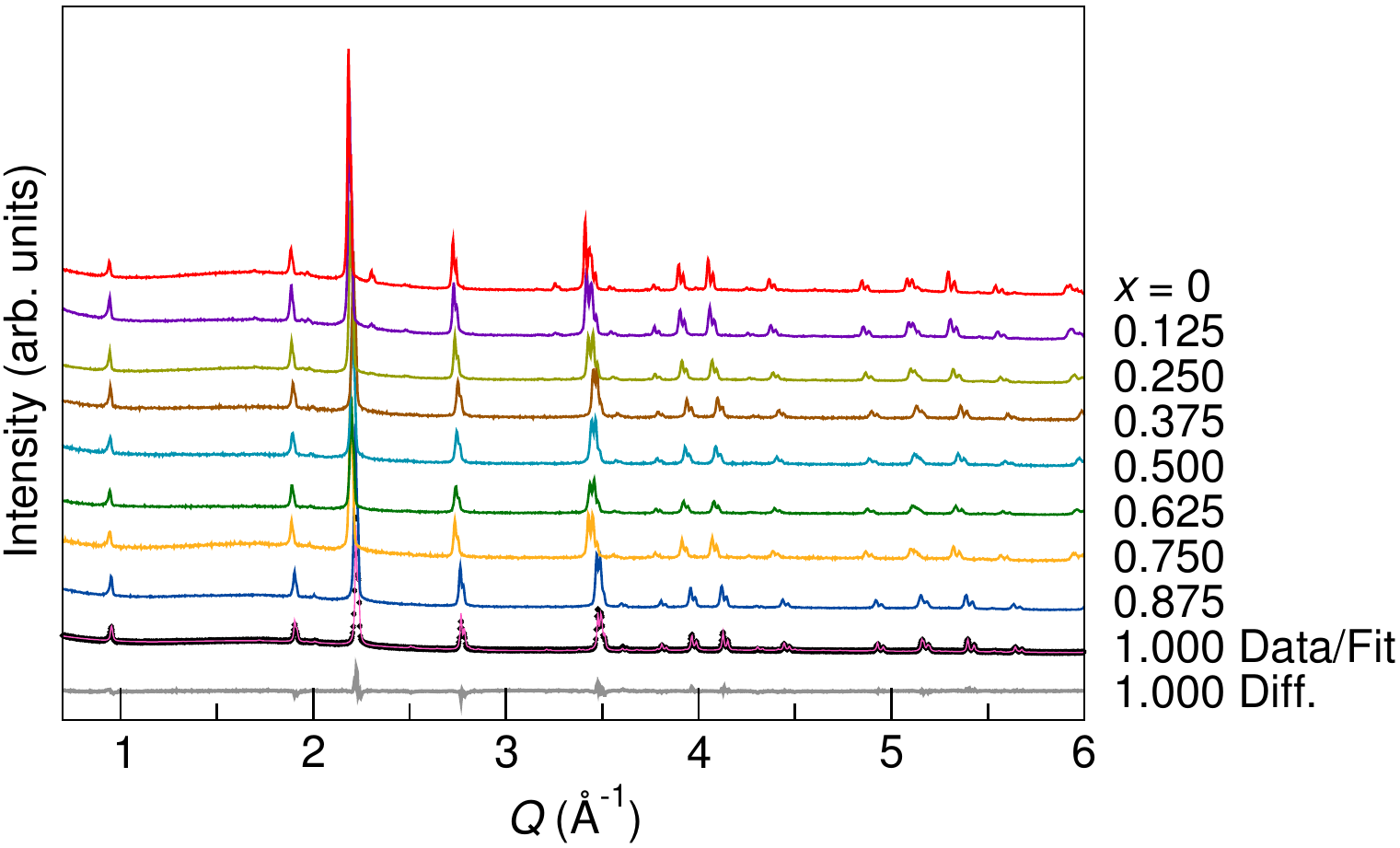}
	\caption{Room-temperature X-ray diffraction patterns of \mf\ show consistent formation of the same structure type, without impurities, and with consistent peak width. The refinement to the \f\ end member with the difference curve is shown.
	}
	\label{fig:xrd}
\end{figure} 

\begin{figure}
	\centering\includegraphics[width=0.95\columnwidth]{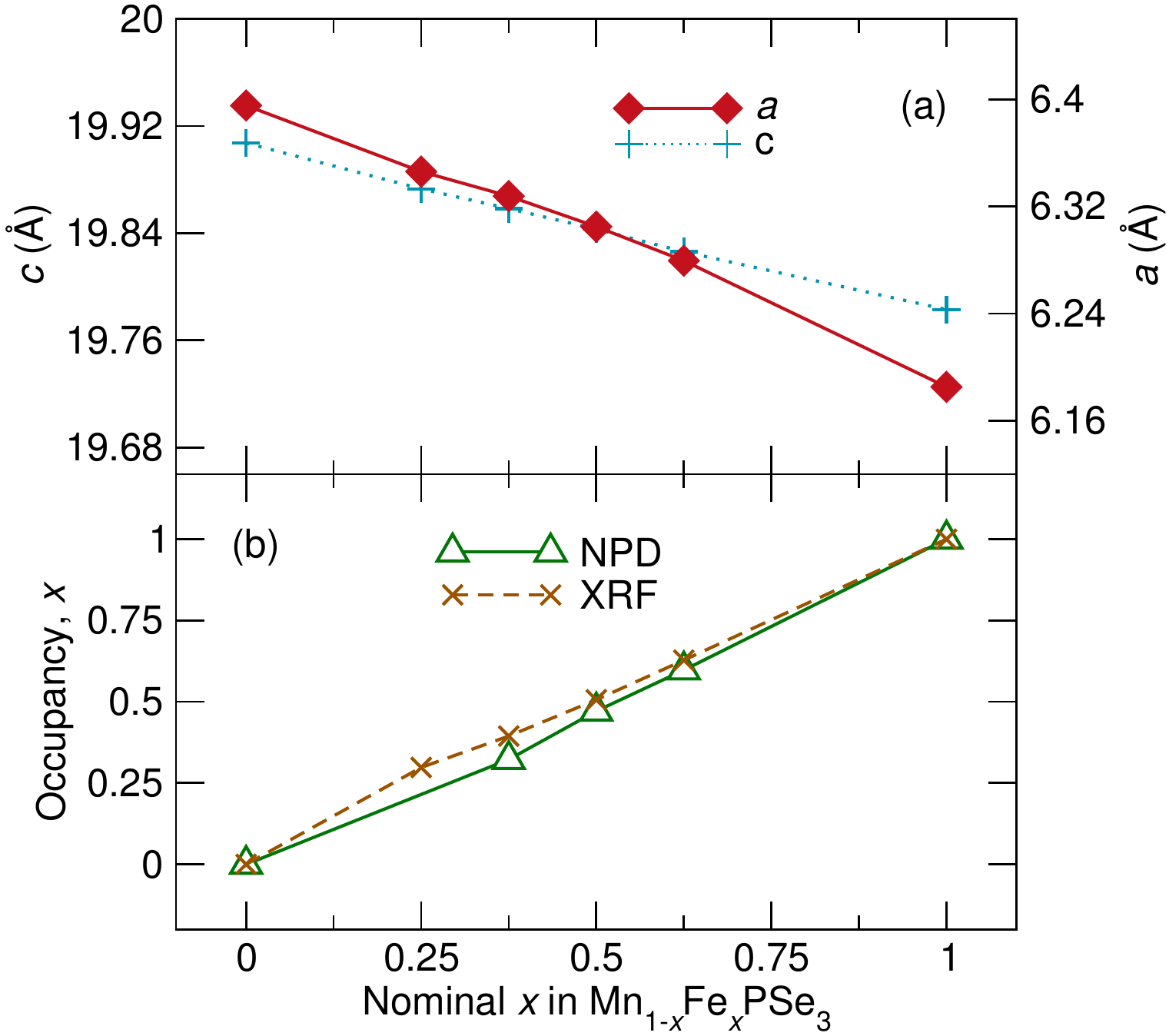}
	\caption{Lattice parameters (a) refined from neutron diffraction data show linear variations from Mn/Fe substitution, {\color{black} with $R\overline{3}$ space group}. In (b), agreement within 5\% is seen in the neutron-refined Mn/Fe occupancies and the Mn/Fe ratio obtained from XRF. Taken together, the data indicate a random solid solution. Error bars are smaller than symbols in all cases.
	}
	\label{fig:xrf-latparam}
\end{figure}

Laboratory powder X-ray diffraction patterns for all compositions in \mf\ at room temperature are shown in Figure \ref{fig:xrd}. The Rietveld refinements for the diffraction patterns indicate that all synthesized compositions are phase pure. Due to the long annealing times (30 days) and the consistent peak width of reflections at high $Q$, it is apparent that the cation ordering is random and relaxed. However, the occupancies of Mn and Fe are indistinguishable by X-ray diffraction analysis and were refined separately by neutron diffraction. 
The Mn/Fe ratios obtained from XRF data are plotted in Figure \ref{fig:xrf-latparam} and slightly overestimate the Fe content by less than 10\%. The XRD-refined chemical contraction of the unit cell from \m\ to \f\ varies smoothly, with a total
change of about 4\% in $a$ and 2\% in $c$. This provides a consistent picture that the individual samples are truly a solid solution.

\begin{figure}
	\centering\includegraphics[width=0.9\columnwidth]{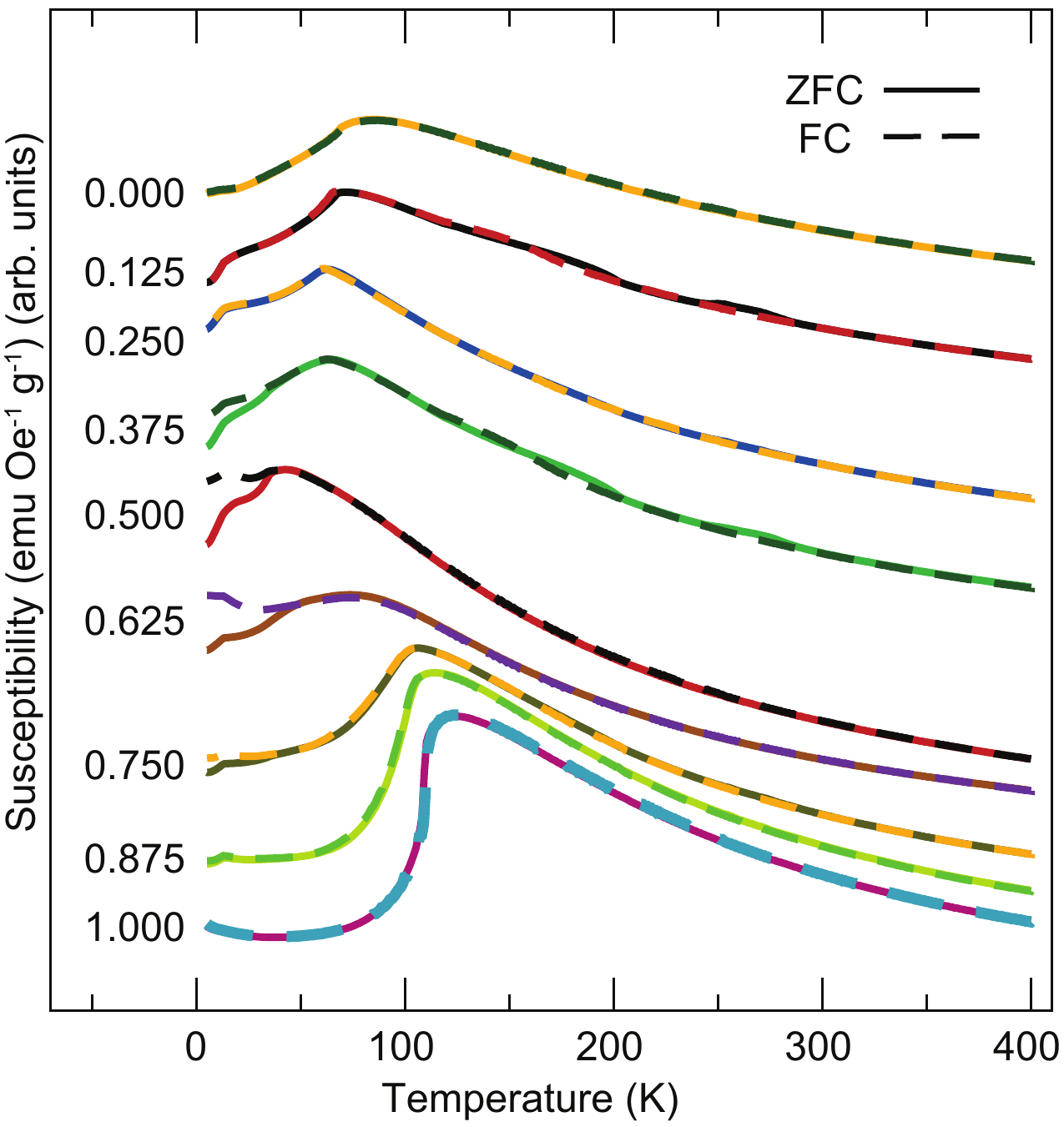}
	\caption{Magnetic susceptibility under zero-field cooling and field cooling with $H = 100$~Oe for all samples in the \mf\ range. The most apparent proxy for Neel temperature is the maximum in susceptibility $T_{\textrm{max}}$, evident for each curve. Only samples from $x = 0.375$ to $0.625$ show irreversibility, as evidenced in deviation of the ZFC and FC susceptibilities. 
	}
	\label{fig:mag-susc} 
\end{figure} 

\begin{figure}
	\centering\includegraphics[width=0.95\columnwidth]{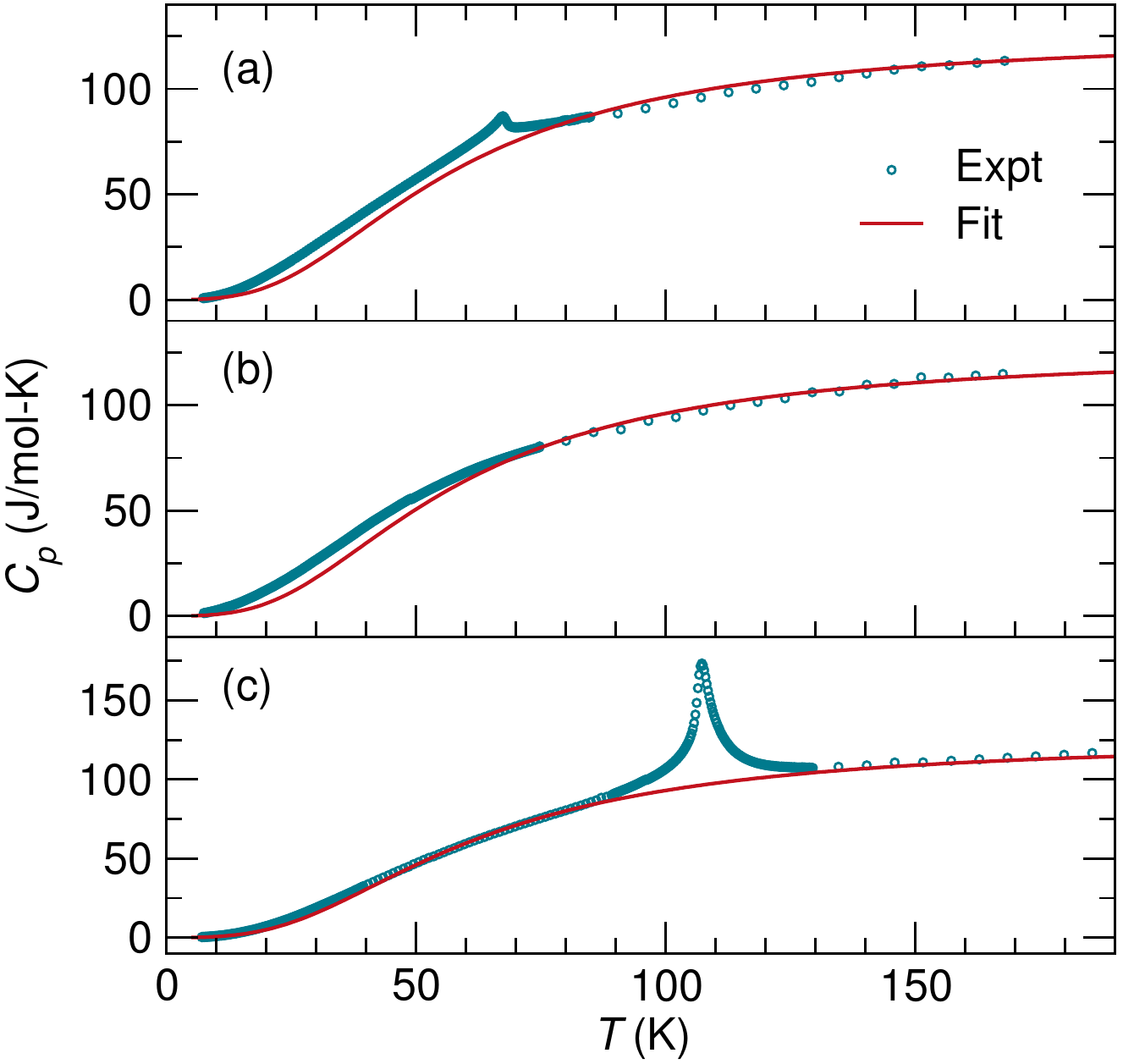}
	\caption{{\color{black}Heat capacity of the end members (a) \m\ and (c) \f\ display
		 clear peaks at the first-order $T_N$. The peak in \m\ is weaker due to the lack of orbital contribution when $S=5/2$.
		At $x$ = 0.5 in (b), the transition is broadened due to slow growth of nano-sized competing magnetic domains, but the total contribution can still be extracted from the Debye fits.}
	}
	\label{fig:cp} 
\end{figure}

Magnetic susceptibility measurements for all compositions in \mf\ are shown in Figure \ref{fig:mag-susc}.
For low-dimensional systems, the value of $T_N$ as measured by specific
heat is not always directly correlated to the maximum in the susceptibility
versus $T$, and a broad maximum above $T_N$ is caused by
short-range spin correlations.\cite{joy1992magnetism,yusuf2010two,bera2017zigzag,lynn19892d}
Here the $T_N$ from heat capacity (Figure \ref{fig:cp})
is more closely tracked by the point where the slope of the $\chi-T$ curve is maximized.
The heat capacity of the $x = 0.5$ sample shows no lambda anomaly,
although the general features of the susceptibility vary smoothly with $x$.

Curie-Weiss temperatures $\theta$ and effective magnetic moments
($\mu_{\textrm{eff}}$) were extracted from the susceptibility over the 280-400~K temperature range.
The $\theta$ values are negative and summarized in Table \ref{tab:mag}, indicating short-range antiferromagnetic interactions in all compositions, and quite strong $\theta = -146~$K in \m, which gradually weakens with Fe substitution.
The effective magnetic moments $\mu_\textrm{eff}$ of \m\ 
($5.9\mu$$_B$) and \f\ ($5.2\mu$$_B$) indicate that both 
\Mn\ and \Fe\ are present in a high-spin state with $S = 5/2$ 
and $S = 2$. The $\mu_{\textrm{eff}}$ off all compounds agree
roughly with the ideal values, except for the $x=0.875$ and $x=1$ samples,
where $T_\textrm{max}$ is sufficiently high that strict
adherence to  Curie-Weiss behavior is not expected below 400~K.

\begin{table*}
	\caption{\label{tab:mag} 
		Expected values and measured parameters from magnetic susceptibility measurements and fits to Curie-Weiss behavior ($\mu_\textrm{eff}$ and $\theta$). 
	}
	\centering
	\begin{tabular}{p{3cm}p{2cm}p{3cm}p{3cm}p{2.1cm}p{2.1cm}p{2.1cm}}
		\hline\hline
		$x$ in \mf\   & $S_{ideal}$ & $\mu_\textrm{eff,ideal}$ ($\mu_B$) & $\mu_\textrm{eff,expt}$ ($\mu_B$) & $\theta$ (K) & $T_\textrm{max}$ (K) & $T_\textrm{split}$ (K)  \\
		\hline 
		0.000  &   5/2  &   5.92 &  5.90 &   -146 &    84 &  -  \\
		0.125  &   2.44 &   5.79 &  5.98 &   -150 &   70 &  -  \\
		0.250  &   2.38 &   5.66 &  5.98 &   -130 &   61 &  -  \\
		0.375  &   2.31 &   5.54 &  5.68 &   -97.7 &  63 &  40  \\
		0.500  &   2.25 &   5.41 &  5.76 &  -88.6 &   40 & 40 \\
		0.625  &   2.19 &   5.28 &  4.82 &  -56.6 &   73 & 46 \\
		0.750  &   2.13 &   5.15 &  4.93 &  -39.7 &   105 &  43 \\
		0.875  &   2.06 &   5.03 &  5.43 &  -28.3 &   113 &  -  \\
		1.000  &   2.00 &   4.90 &  5.24 &  -8.86 &   124 &    -  \\
		\hline \hline
	\end{tabular}
	~\\
\end{table*}

%
%
%
%

Splitting between the ZFC and FC susceptibilities in Figure \ref{fig:mag-susc} is only observed from $x = 0.375$ to $x = 0.75$ and occurs around $40~K$. The onset of this irreversibility is denoted $T_\textrm{split}$ in Table \ref{tab:mag} and suggests uncompensated spins that arise at boundaries of domains with dissimilar magnetic orderings, so it
is not evident in the end members. The uncompensated surface spins of the domains can behave in a glassy or disordered way. The highest degree of irreversibility is observed as $x$ approaches $0.5$ suggesting a higher uncompensated surface contribution form magnetic domains in intermediate compositions.



The total heat capacity measurements in Figure \ref{fig:cp} only display an 
obvious $\lambda$ anomaly for the end members \m\ and \f, but
even fitting the $x=0.5$ sample to the Debye model 
reveals a gradual onset of magnetic ordering.
The large peak in FePSe$_3$ (compared to \m) can be explained by the
magnetoelastic contribution from spin-orbit coupling,
as was suggested for FePS$_3$.\cite{jernberg1984feps3}
Furthermore, the magnetic frustration as viewed by a larger Curie-Weiss $\theta$
versus the susceiptibltiy $T_\textrm{max}$ indicates that \m\ is frustrated,
and slowly orders with increasing domain size upon cooling. This is
reflected in the deviation of $C_p$ versus the Debye fit in Figure \ref{fig:cp}(a).

The total heat capacity at low temperatures is a combination of electronic, 
lattice and magnetic contributions 
$C_{total} = C_{elec} + C_{lat} + C_{mag}$, where $C_{elec}$ is $\gamma T$, $C_{lat}$ is $\beta T^3 + \alpha T^5$. The fit to the heat capacity at low temperatures ($7-10~K$) was made using $C_{lat}$ since these chalcogenides are insulators with high resistivity of the order of $10^6$~$\Omega$-m to estimate Debye temperatures. 
The high-temperature heat capacity data was then fit using the Debye model to better estimate $C_{lat}$ and Debye temperatures.
$C_{mag}$ was calculated by $C_{total}-C_{lat}$ and $C_{mag}/T$ vs $T$ plot was integrated to give the entropy associated with the magnetic transition. 
The theoretical limit to the statistical magnetic entropy for complete ordering of \Mn\ ($S=5/2$) should be $R~ln(2S+1) = 14.89~J~mol^{-1}~K^{-1}$ and of \Fe\ ($S = 2$) should be $13.38~J~mol^{-1}~K^{-1}$.  It is clear from Figure \ref{fig:cp} that
the \mf\ does precisely track Debye-like behavior, as is typical for
similar materials, \cite{pei2016spin} but rough agreement is seen:
The entropy calculated for $x = 0.0, 0.5$ and $1.0$ amount to $13.84~J~mol^{-1}~K^{-1}$, $13.23~J~mol^{-1}~K^{-1}$ and $10.73~J~mol^{-1}~K^{-1}$ with respective Debye temperatures of $235~K$, $240~K$ and $250~K$.
These values indicate that the ordering in intermediate compositions
is still transitioning from states that are nearly fully disordered
to fully ordered over the measured temperature range.


\subsection{Progression of magnetic ordering across the \mf\ compositional range}


%
%
%
%
%
%



\begin{figure}
	\centering\includegraphics[width=0.95\columnwidth]{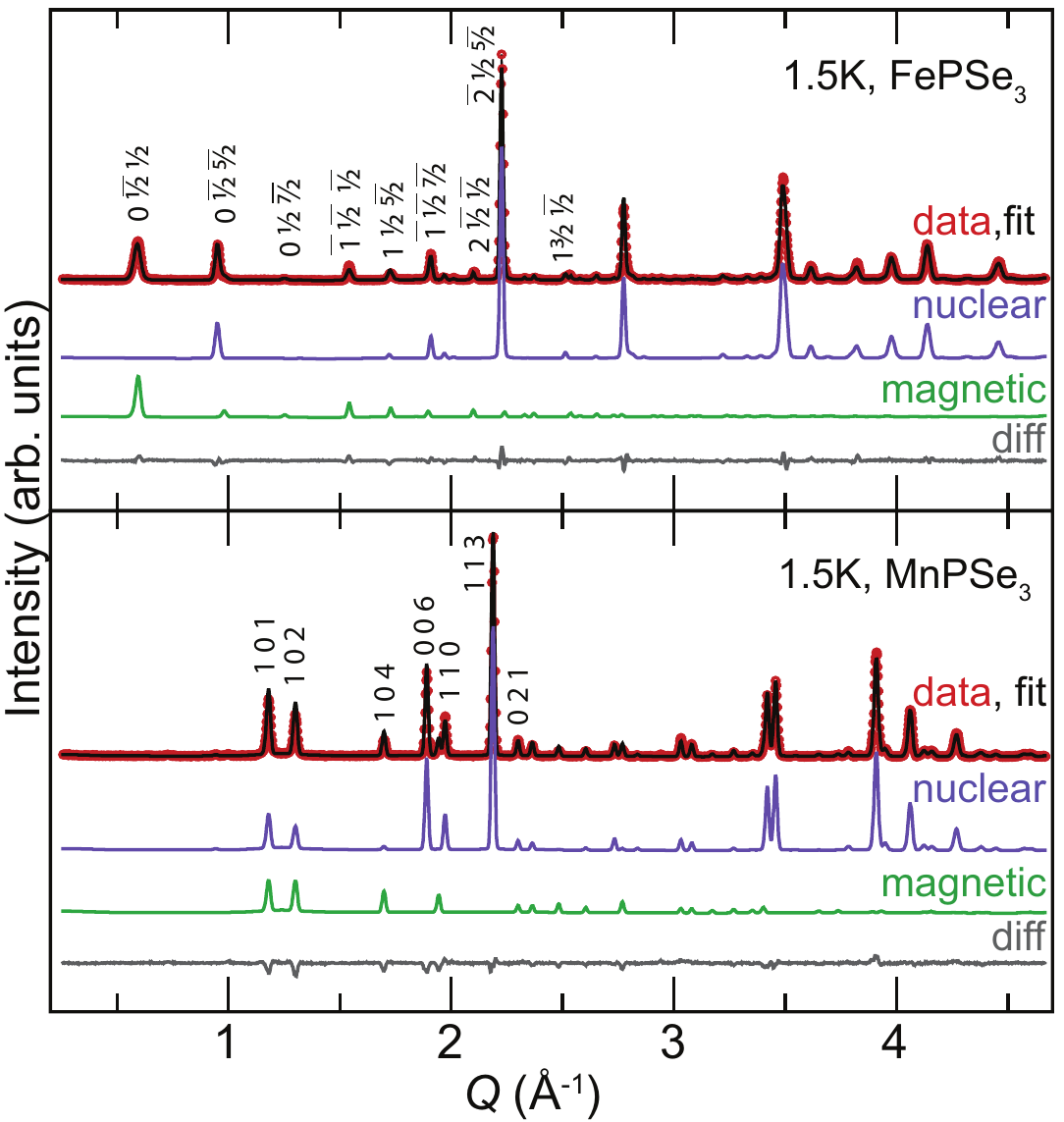}
	\caption{
		Refinements to neutron powder diffraction data at $T = 1.5$~K for \f\ and \m\ show clear signatures from magnetic ordering. All magnetic intensity in \m\ lies on nuclear reflections since $k = [000]$.}
	\label{fig:hb2a-refine} 
\end{figure}

\begin{figure*}
	\centering\includegraphics[width=1.4\columnwidth]{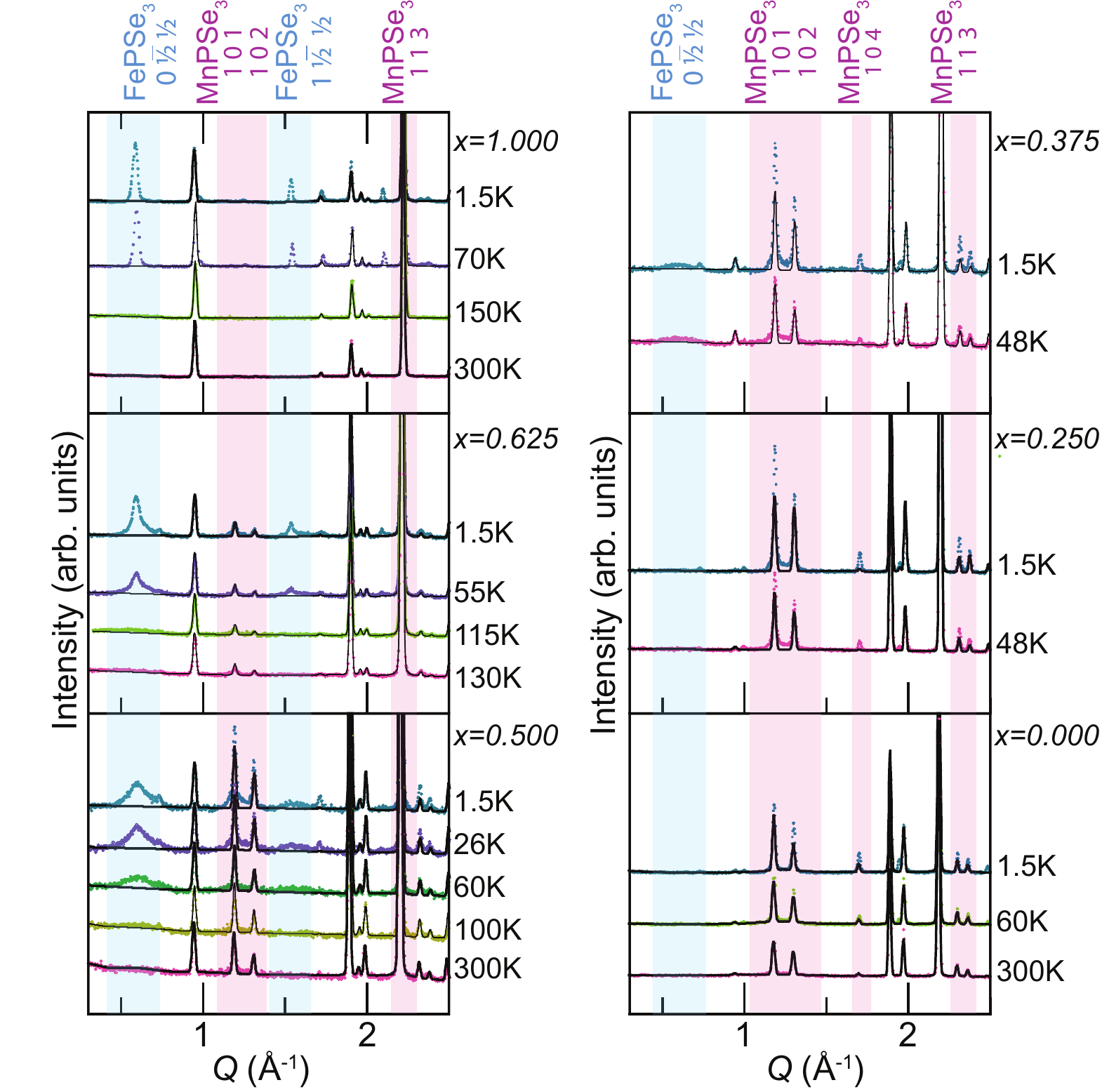}
	\caption{Evolution of magnetic ordering peaks with temperature and composition. The nuclear fits have been shown in black to clearly identify magnetic intensities at various temperatures. Peaks corresponding to \f-type and \m-type magnetic ordering have been highlighted in blue and pink respectively. Presence of broad diffuse magnetic peaks caused by short-range order is seen in intermediate compositions as compared to sharp magnetic peaks in end members.} 
	\label{fig:hb2a-evolution} 
\end{figure*}

Our refined neutron powder diffraction data at $T=1.5$~K is shown for the end members \m\ and \f\ in Figure \ref{fig:hb2a-refine}. We verify the magnetic propagation vectors $k = [000]$ 
and $k=[\frac{1}{2}0\frac{1}{2}]$, respectively. \cite{wiedenmann1981neutron} 
The average magnetic moments on Mn$^{2+}$ and Fe$^{2+}$ in the end members were refined to 3.6~$\mu_B$ and 4.2~$\mu_B$, respectively. 
{\color{black} The in-plane direction of the \Mn\ moments cannot be determined from powder neutron scattering due to the hexagonal $R\overline{3}$ symmetry.}





\begin{figure}
	\centering\includegraphics[width=0.95\columnwidth]{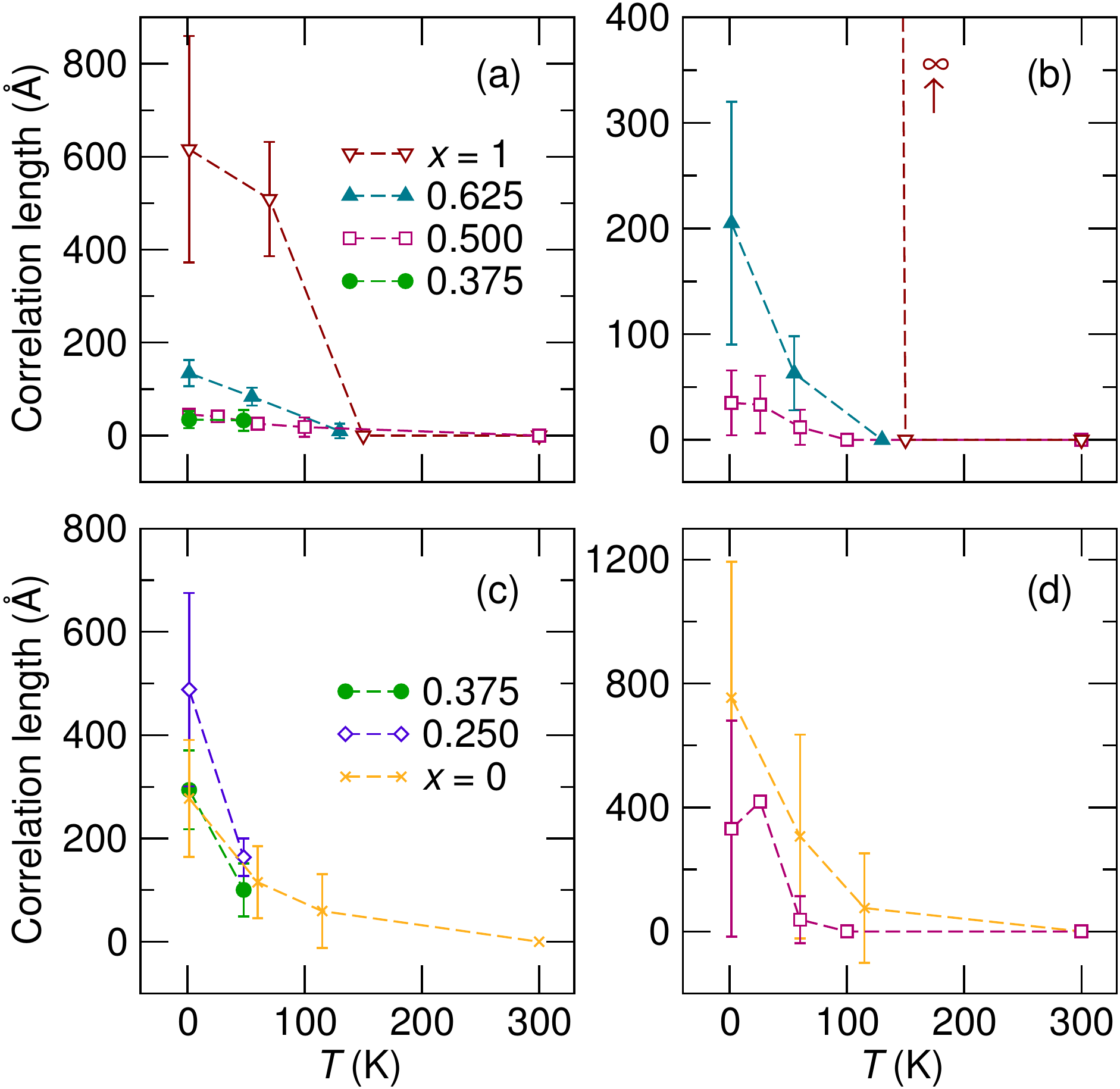}
	\caption{Magnetic correlation lengths for various compositions calculated as a function of temperature for 
		(a) \f-type $0 \bar{\frac{1}{2}} \frac{1}{2}$ at $Q = 0.6~\textrm{\AA}^{-1}$,  
		(b) \f-type $\bar{1} \bar{\frac{1}{2}} \bar{\frac{1}{2}}$ at $Q = 1.53~\textrm{\AA}$$^{-1}$,
		(c) \m-type $101$ at $Q = 1.17~$\AA\, and
		(d) \m-type $102$ at  $Q = 1.30~$\AA.}
	\label{fig:corr-lengths}
\end{figure}

{\color{black} The smaller magnitudes of neutron-refined magnetic moments versus the paramagnetic moments from susceptibility can be attributed to uncertainty in the canting of magnetic moments or to small domain sizes with imperfect magnetic ordering in \m. The magnetic structures of the analogous sulfides remain a topic of active investigation. \cite{lanccon2016magnetic, ressouche2010magnetoelectric} The magnetic structure of MnPS$_3$ was identified with a propagation vector of $k = [000]$ where the \Mn\ moments lie at an angle of {\color{black} 8$^\circ$} from the $c^\star$ axis, as compared to the previously-published magnetic structure where the magnetic moments are along $c^\star$. \cite{ressouche2010magnetoelectric} 
If \m\ also has a canted configuration, Rietveld analysis with \Mn\ moments lying in the $ab$ plane would cause the calculated magnetic moments to be lower than the true value. On the other hand, small magnetic correlation lengths in \m\ are shown in Figure \ref{fig:corr-lengths}, indicating a lack of perfect long-range magnetic ordering.
The prevalence of disordered regions between these domains would also lead to a smaller neutron-refined magnetic moment. 
}

Across the compositional range, a few key changes should be noted
in the neutron diffraction patterns at 1.5~K, shown in Figure \ref{fig:hb2a-evolution}: first,
the magnetic reflections in \f\ are clearly broadened (and although
it is more subtle, there is substantial diffuse scattering from magnetic
intensity in \m), and there is a progression of mixing and broadening
of the magnetic Bragg contributions from both phases as intermediate
values of $x$ are examined.

\begin{figure}
	\centering\includegraphics[width=0.8\columnwidth]{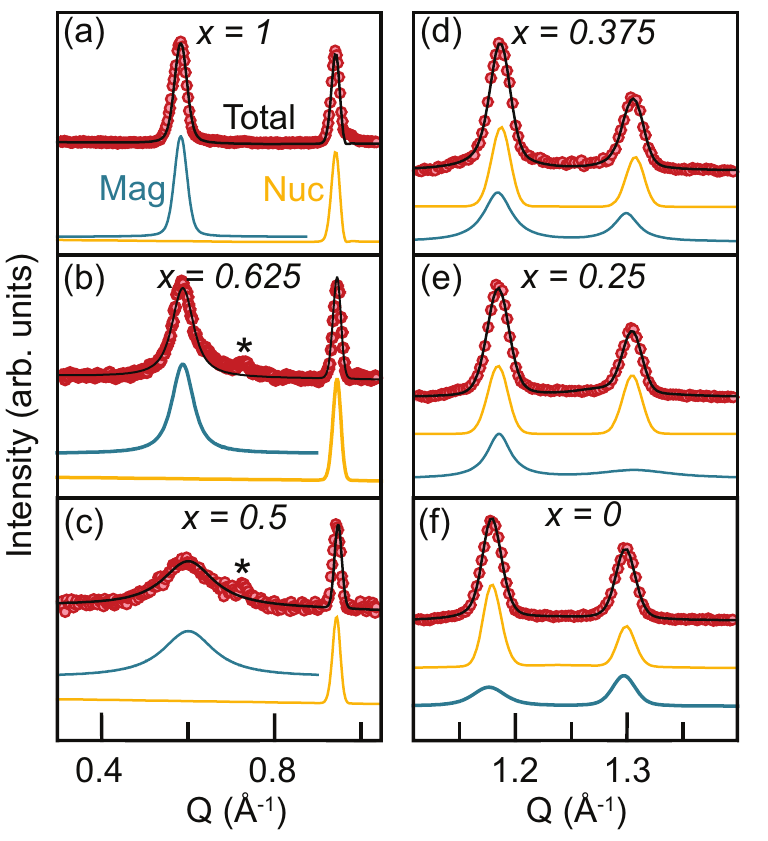}
	\caption{Magnetic and nuclear contributions obtained from Rietveld refinements of neutron diffraction patterns at 1.5K for (a) $x = 1$, (b) 0.625, (c) 0.5, (d) 0.375, (e) 0.25, and (f) 0.
	{\color{black} A minor peak marked with an asterisk is assumed to be a magnetic peak from Fe$_7$S$_8$ that was below the detection level of our XRD data.
	}
}
	\label{fig:lorentz}
\end{figure}

In \f, the broadening of the $0\frac{1}{2}\frac{1}{2}$ magnetic reflection
is not immediately apparent from Figure \ref{fig:hb2a-refine}, but
upon closer inspection in Figure \ref{fig:lorentz},
it is significant and can be refined as a Voigt contribution corresponding
to a correlation length $L=600\pm 200$~\AA,
 and remains broad at $T=70$~K to
$L=500\pm 100$~\AA. 
This peak broadens further into a diffuse, but still detectable, contribution
at 150~K, which is higher than $T_N=124$~K for \f,
indicating short-range magnetic correlations that are common
for low-dimensional materials. \cite{yusuf2010two,bera2017zigzag,lynn19892d}
For a higher-angle $\bar{1} \bar{\frac{1}{2}} \bar{\frac{1}{2}}$ magnetic peak,
the correlation lengths are not determinable within the limits of instrumental and sample broadening.

In other magnetic compounds with strong crystalline anisotropy
such as such as Sr$_2$YRuO$_6$\cite{granado2013two}, CrTa$_2$O$_6$\cite{saes1998structure} and La$_2$O$_3$Mn$_2$Se$_2$.\cite{ni2010physical}, magnetic domains that exhibit strong
correlations in two dimensions above 3D long range magnetic transition temperature are typically 
modeled by Warren-type peaks,\cite{warren1941x} which are characterized by long tails
with increasing $Q$, similar to turbostratic nuclear disorder in
layered compounds and clays.
While the layered structure of \mf\ could play host to such disorder,
we observe neither nuclear disorder nor Warren-type tails on the magnetic
peaks. Instead, the magnetic peaks are best described as Lorentzian contributions after instrumental and crystallite size corrections (Figure \ref{fig:lorentz}). 
This implies that the short range ordering present in \mf\ has a significant interplane component, unlike other 2D materials such as Sr$_2$YRuO$_6$, CrTa$_2$O$_6$ and La$_2$O$_3$Mn$_2$Se$_2$. This behavior is corroborated by the fact that the broad magnetic peaks correspond to $hkl$ family of planes, instead of $hk0$.

For samples where $0.675 \leq x \leq 0.375$, magnetic peaks
are broadened and the two $k$-vectors coexist. The extracted
correlation lengths for these with varying composition 
and temperature are plotted in Figure \ref{fig:corr-lengths}.
Interestingly, only the \f\ end member at $x=1$ shows
domain sizes that are large enough that the peaks are
not broadened versus the nuclear peaks. 
Correlation lengths drop more steeply for \f-type ordering as compared to \m-type ordering for intermediate compositions.  
{\color{black} This} could be explained by stronger anisotropic and hence less susceptible character of \m\ as compared to \f.

\subsection{Nature of and driving forces for the coexistence of magnetic domains}

It is clear from the susceptibility and diffraction measurements
that \mf\ exhibit mixed magnetic ordering below $T_N$.
The layers containing magnetic cations are separated by a van der Waals gap 
on the order of $\sim7$~\AA, which prohibits
direct exchange and superexchange interactions between layers. 
The intralayer neighboring magnetic interactions are much stronger,
as evidenced by the non-Curie-Weiss behavior and diffuse magnetic
scattering above $T_N$. 
Clearly, the differences between this system and other mixed magnets
(which typically result in spin glasses) should
be understood.
For a random cation mixture on \mf, a Hamiltonian for the spin
interactions can be written: 
\begin{equation}
H = -2 J_{Mn} - 2 J_{Fe}- 2 J_{MnFe} - D_{Mn} - D_{Fe} \, ,
\end{equation}
where, 
\begin{eqnarray}
J_{Mn} &=& J_{MnMn}{\sum_{<i,j>}}\vec{S}_{Mn_{i}}\cdot\vec{S}_{Mn_{j}} \, , \nn \\
J_{Fe}&=&J_{FeFe}{\sum_{<k,l>}}\vec{S}_{Fe_{k}}\cdot\vec{S}_{Fe_{l}} \, ,\nn \\ 
J_{MnFe}&=&J_{MnFe}{\sum_{<p,q>}}\vec{S}_{Mn_{p}}\cdot\vec{S}_{Fe_{q}} \, ,\nn \\
D_{Mn}&=&D_{Mn}{\sum_{i}}{{(S_{Mn_{i}}^{z}})^2} \, , \nn \\
D_{Fe} &=& D_{Fe}{\sum_{k}}{{(S_{Fe_{k}}^{z}})^2}
\end{eqnarray}



%
%


Here, $J$ are exchange interactions between two neighboring magnetic ions
and $D$ denotes the anisotropy. 
$D_\mathrm{Mn} < 0$ and $D_\mathrm{Fe} > 0$ for \m\ and \f\ as per their Heisenberg
and Ising nature, respectively. 
\m\ is highly anisotropic as determined by single-crystal magnetic 
susceptibility measurements carried out by Jeevanandam \cite{jeevanandam1999magnetism} with a single-ion exchange anisotropy $D = 26.6$~K, 
which is approximately five times the exchange interaction ($-5.29~K$).
No comparable susceptibility measurement exists for \f\ to estimate 
the value of $D$. However, the exchange interaction $J_\mathrm{FeFe}$ 
is of similar magnitude (between 3.7 and 10.4~K) to that of \m\, 
but ferromagnetic as determined by Wiedenmann.\cite{wiedenmann1981neutron} 

{\color{black}At first glance, it may seem surprising that $D_\mathrm{Mn}$ is large, given the 3$d^5$ electron configuration and zero orbital contribution. 
Magnetic anisotropy of Mn$^{2+}$ compounds is perhaps best understood in the context of the Mn$X_2$ halides where $X$ = (F, Cl, Br, I). 
For the larger anions, covalency increases along with the ligand contribution to spin-orbit coupling.
This increase in covalency, coupled with the highly anisotropic crystal structures of the halides (and the selenophosphates we investigate here) can be most dramatically observed in the magnetic anisotropy and in the strong photoluminescence and magnetic dichroism of MnI$_2$.\cite{hoekstra_optical_1983,ronda_photoluminescence_1987} 
MnI$_2$ has the Cd(OH)$_2$ structure type, with Mn in slightly trigonally-distorted octahedra, like \m, and without ligand covalency the observed optical transitions would be forbidden.
A similar line of reasoning explains single-site anisotropy in Mn$^{2+}$ single-molecule magnets\cite{chowdhury_heavy_2017} and the anisotropy in CrI$_3$, which is also layered with a 3$d^3$ ground state that possess magnetic anisotropy due to spin-orbit coupling.\cite{lado_origin_2017}
Interplane ordering is more likely dipolar in nature.\cite{lhotel_subtle_2007,sato_successive_1995}
The treatment of spin-orbit-driven anisotropy in MnPSe$_3$ in particular  has been laid out by Jeevanandam and Vasuvedan.\cite{jeevanandam1999magnetism} Covalency and the spin-orbit coupling are both higher for selenium (1463 cm$^{-1}$) than for sulfur (1463 cm$^{-1}$),\cite{barnes_electric_1954} which in turn has a substantial effect on zero-field splitting parameter $D$. A more precise decomposition of the effects that lead to anisotropy in the chalcophosphates remains to be conducted, as the polyanionic species (P$_2$Se$_6^{4-}$) are not equivalent to selenides.
}

%

Assuming similar magnitudes of $D_\mathrm{Fe}$ and $D_\mathrm{Mn}$, the question
is what ordered states are accessible by a random 2D-sheet
mixture of these cations.
Fishman and Aharony have provided theoretical models for random alloys of two antiferromagnets with different periodicities, different anisotropies and different interactions in separate studies, \cite{fishman1978phase,fishman1979phase,fishman1980phase} but their results cannot be directly applied to our system 
which represents a combination of all three forms of competition. 

A solid solution of analogous sulfides, 
on the other hand, results in a spin glass state at intermediate
compositions.\cite{takano2003magnetic} 
Both MnPS$_3$ and FePS$_3$ order antiferromagnetically with spins 
normal to {color{red} the} \textit{ab} plane and $k = [000]$ and 
$k =  [01\frac{1}{2}]$, respectively. 
In MnPS$_3$, each \Mn\ is antiferromagnetically coupled 
with its nearest neighbors in the plane and there is ferromagnetic 
coupling between the planes. In FePS$_3$, each \Fe\ is ferromagnetically coupled with two nearest neighbors and antiferromagnetically with the third one and forms zigzag spin chains coupled antiferromagnetically within each layer. MnPS$_3$ is magnetically isotropic with a very {\color{black}small} $D = 0.105$~K, with exchange interactions of $J_1 = -9.1$~K, $J_2 = -0.83$~K and $J_3 = -2.15$~K. \cite{wildes1998spin} The nature of small anisotropy is debated between dipolar anisotropy and single ion anisotropy, however {\color{black} only its magnitude is relevant to our comparison}. FePS$_3$, on the other hand, is anisotropic with $D~=~31.7$~K, approximately double the exchange parameters: $J_1 = 17.4$~K, $J_2 = -0.48$~K, $J_3 = -11.4$~K. \cite{lancon2016magnetic}  
{\color{black} The sulfides form a spin glass when mixed randomly because competing antiferromagnetic and ferromagnetic exchange interactions within the planes are frozen without long-range preference for specific orientations (small $D$).\cite{masubuchi2008phase, takano2003magnetic}
The local \Mn\ symmetries of $M$PS$_3$ and $M$PSe$_3$ compounds both contain trigonally-distorted octahedra, with deviations about 3-4$^\circ$ and 5$^\circ$, respectively, and short/long bond distances of 2.70/2.74 \AA\ and 2.62/2.63 \AA, respectively.\cite{ouvrard_structural_1985,wiedenmann1981neutron} 
Formally, the site symmetry is actually higher for the selenide ($C_3$ versus $C_2$) as a consequence of the interlayer stacking. Small differences in local symmetry are not expected to dominate magnetic anisotropy, although systematic theoretical and computational work could shed additional light on the magnitude of these effects.
}

{\color{black} In contrast to the sulfide analogs, the absence of a spin glass 
state in \mf\ can be explained by the dominance of anisotropies $D_\mathrm{Mn}$ and $D_\mathrm{Fe}$ 
over the exchange interactions.
 }
The tendency to obey a particular magnetic ordering increases with increasing anisotropy. Even small local chemical clustering in a randomly mixed solid solution can change the spin dynamics and segregate the system into coexisting magnetic domains of the favored end members. Local regions rich in \Mn\- or \Fe\-type ions can continue to polarize the magnetic ordering in their vicinity resulting in a two-phase competition region between $x = 0.25$ and $x = 0.875$. 


\begin{figure}
	\centering\includegraphics[width=1\columnwidth]{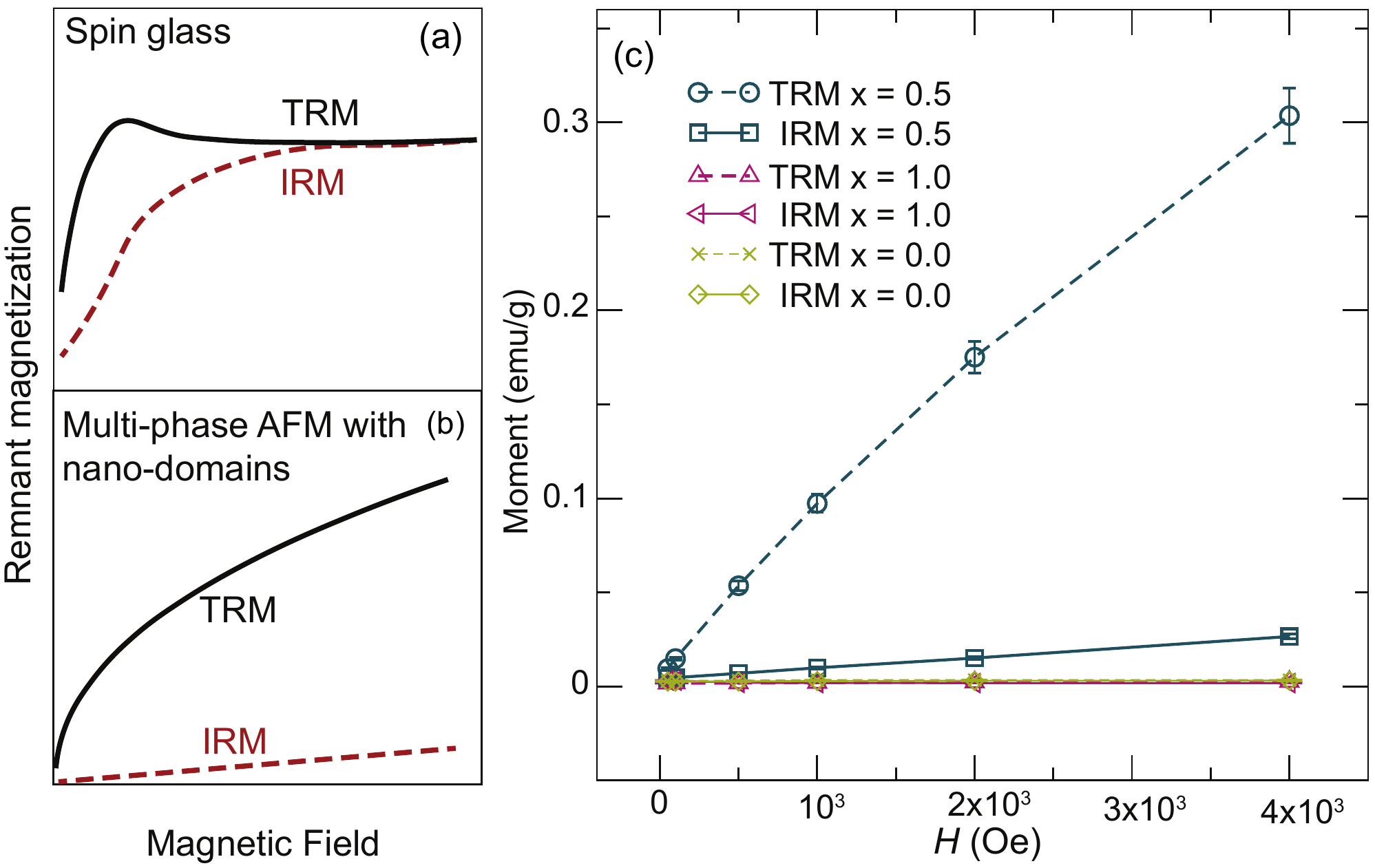} \\
	
	\caption{Schematic figure of thermoremanent magnetization (TRM) and isothermal remanent magnetization (IRM) for (a) spin glass, (b) nano-wires, adapted from article by article by Benitez et. al.\cite{benitez2011fingerprinting}, (c) Thermoremanent magnetization (TRM) of $x$ = 0.5 samples shows
		an sub-exponential increase with field and divergence from isothermal magnetization (IRM),
		typical of an antiferromagnetic system with small domains and polarizable domain walls. The
		end members \m\ and \f\ show no remanence.}
	\label{fig:trm-irm} 
\end{figure}

%
%

%
Among the SG and 2-phase models that are possible ground
states for such randomly-mixed 2D systems, each has its own
tendency for formation based on $J$ and $D$ competition.
The macroscopic response of these scenarios manifest in changes
in the amount of uncompensated spins and their time-dependent
susceptibility. Clearly, the spin glass
scenario is ruled out of \mf\ due to the high amount
of ordered moment observed in the neutron diffraction data, 
but additional confirmation can be seen in time-dependent
magnetization measurements.

{\color{black} Thermoremanant magnetization (TRM) and isothermal remanent magnetization (IRM)} curves for ideal bulk antiferromagnets should be 
zero,\cite{rodriguez2017producing} and 
higher values of TRM versus IRM denote irreversibility 
as embodied in a spin-glass (evenly-distributed frozen spins) or nano-domain behavior with a large fraction of uncompensated surfaces, occasionally seen in core-shell nanoparticles.
Both behaviors are shown schematically in Figure \ref{fig:trm-irm}. \cite{benitez2011fingerprinting}
For a spin glass, the IRM increases with increasing field, then meets the TRM curve at
moderate field values, where both then saturate. The TRM also exhibits a characteristic peak at intermediate fields. TRM-IRM curves for antiferromagnetic nanoparticles have been measured and show an increasing TRM and IRM with no signs of saturation, a behavior that has been often compared to a 2D-DAFF response. \cite{benitez2008evidence}


{\color{black}The TRM and IRM measurements}
at 5~K on \mf\ for $x = 0.0, 0.5, 1.0$ are shown in Figure \ref{fig:trm-irm}. TRM and IRM for $x = 0.0$ and $1.0$ are negligible (ideal bulk antiferromagnets) as compared to those for $x = 0.5$. For $x = 0.5$, the IRM increases nearly 
linearly, but at a slower rate than TRM.
TRM and IRM for $x = 0.5$ does not saturate at high magnetic 
fields and does not display a spin-glass behavior, but instead matches interface-dominated behavior, which is
seen in systems with small magnetic domain sizes, for example
in Co$_3$O$_4$ nanowires, where uncompensated surface spins lead to irreversibility in addition to the regular antiferromagnetic contribution from the wires.\cite{benitez2008evidence} The decrease in correlation lengths of coexisting clusters of \m\ and \f\ type ordering at intermediate compositions lead to more ``uncompensated surfaces" with random ordering,
 which results in an increasing TRM and IRM. 

\begin{figure}
	\centering\includegraphics[width=0.95\columnwidth]{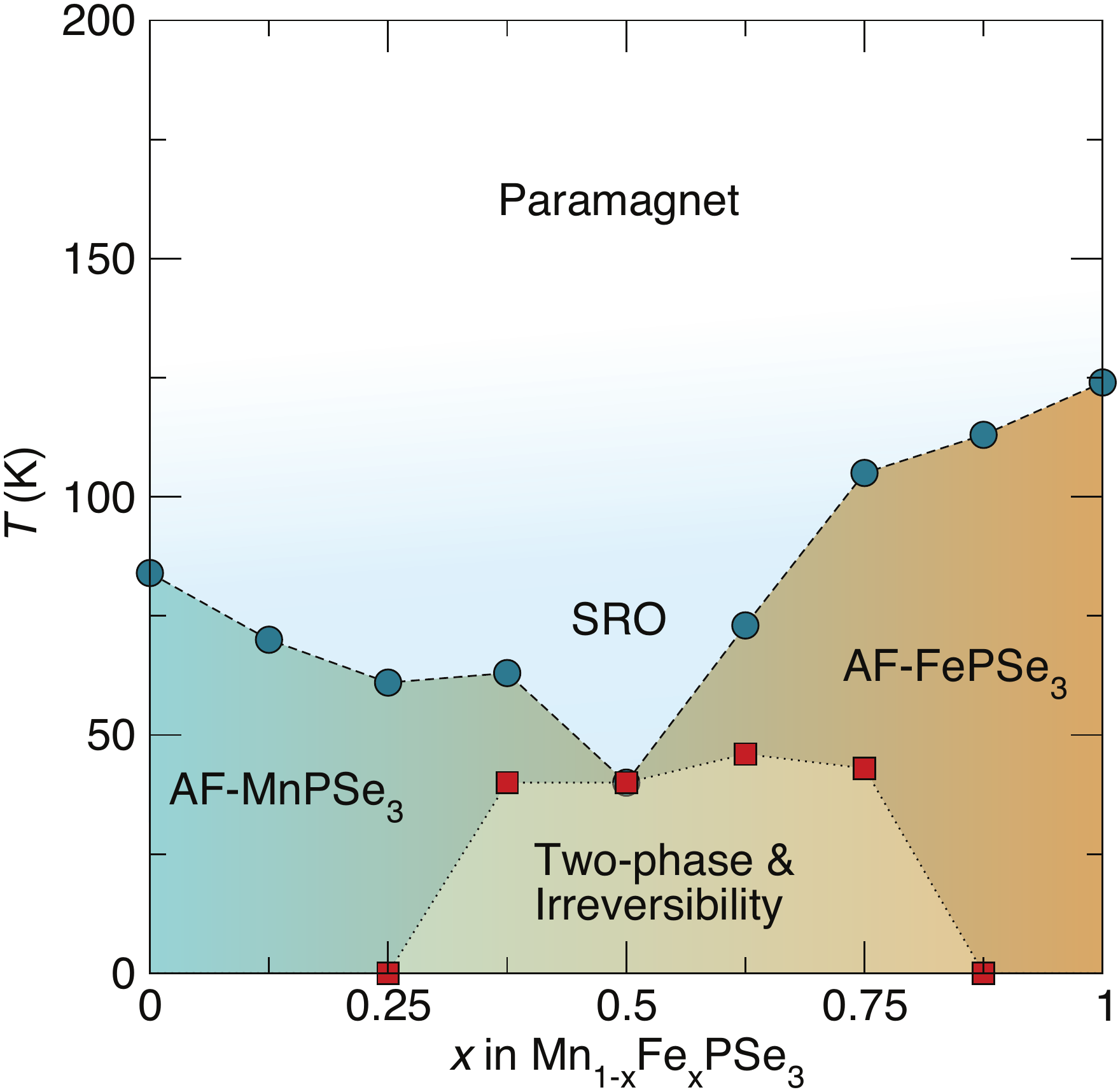}
	\caption{Magnetic phase diagram of \mf\ showing three regions with \m-type, mixed-type and \f-type AFM ordering. The circles represent $T_{max}$ from $\chi-T$ measurements and a crossover from paramagnetic state to a magnetic state, while the two-phase competition region is best denoted by the susceptibility $T_\textrm{split}$ (squares). Short-range order (SRO) is evident from deviation from Curie-Weiss susceptibility and diffuse magnetic nuclear scattering intensity.
	}
	\label{fig:phase-diagram}
\end{figure}

The final magnetic phase diagram of \mf\ is shown in Figure \ref{fig:phase-diagram}. The phase transition lines were drawn based on $T_{max}$ obtained from $\chi-T$ measurements. Between $x = 0.0$ and $x = 0.25$, \m\ type magnetic ordering is present with introduction of short range correlations as $x$ or \Fe\ concentration is increased. $T_{max}$ decreases as $x$ increases and is minimum for $x = 0.5$. Between $x = 0.25$ and $x = 0.875$, mixed ordering or coexistence of \Mn- and \Fe-type ordering is present. The mixed phase forms nano-sized chemically disordered clusters which display both kinds of ordering. The uncompensated surfaces between the clusters increase as the cluster size decreases and the effect can be seen in TRM-IRM, ZFC-FC magnetization and neutron diffraction measurements. Cluster size decreases as a function of chemical disorder present and is smallest for $x = 0.5$. The absence of Schottky anomaly in heat capacity for $x = 0.5$ suggests short range ordering where the transition lines in the phase diagram defined by $T_{max}$ over intermediate compositions are not smooth and very well defined. For $x > 0.875$, \f\ type magnetic ordering is present. The strong dependence of correlation lengths on the \Fe\ concentration for $x > 0.5$ suggests a lower value of anisotropy $D_{Fe}$ as compared to $D_{Mn}$. This is also supported by weak dependence of correlation lengths on \Fe\ concentration for $x < 0.5$.

\section{Conclusions}\label{summary}

In conclusion, we have established a magnetic phase diagram of a mixed spin, mixed interaction, mixed anisotropy and mixed periodicity system \mf\ using a combination of X-ray diffraction, X-ray Fluoroscence, neutron diffraction, DC magnetic susceptibility, TRM, IRM and heat capacity measurements on bulk powder samples. This is the first solid solution study of a random magnet system in metal selenophosphates family. Both kinds of \m\ and \f\ type ordering are found to co-exist at intermediate compositions in the form of nanosized clusters. \f\ type ordering is found to be more susceptible to doping as compared to the \m\ type ordering. A long range ordering does not take place in intermediate compositions upto $1.5~K$ and the broad diffuse scattering peaks are observed in neutron diffraction patterns. The magnetic ordering in intermediate compositions take place over a wide temperature range and does not display a characteristic lambda anomaly in heat capacity. The uncompensated surface spins increase with shorter correlation lengths and are evident in DC magnetization and TRM-IRM measurements. The mixed ordering can be explained by high values of $D$ arising from ligand spin-orbit contributions. Future measurements involving single crystal neutron diffraction can be employed to establish the direction of moments withing the basal plane in \m\. Magnetic domain imaging such as lorentz microscopy and magnetic force microscopy can be used to further characterize and image the anisotropic nature of the domains.


\section*{Acknowledgments}

We acknowledge support from the Center for Emergent Superconductivity, an Energy Frontier Research Center funded by the U.S. Department of Energy, Office of Science, Office of Basic Energy Sciences under Award Number DEAC0298CH1088. Magnetic and heat capacity measurements were carried out in part in the Materials Research Laboratory Central Research Facilities, University of Illinois. 
Neutron powder diffraction measurements conducted at ORNL's High Flux Isotope Reactor was sponsored by the Scientific User Facilities Division, Office of Basic Energy Sciences, US Department of Energy.

\bibliography{mpse3}

\end{document}